\documentstyle[twocolumn,prl,aps]{revtex} 
 
\begin{document} 
\twocolumn[\hsize\textwidth\columnwidth\hsize\csname @twocolumnfalse\endcsname
\draft 
\preprint{} 
\title{ Disappearance of integer quantum Hall effect} 
\author{D.N. Sheng, Z. Y. Weng} 
\address{Texas Center for Superconductivity and Department of Physics\\ 
University of Houston, Houston, TX 77204-5506 }  
\maketitle 
\date{tody}
\begin{abstract} 
The disappearance of integer quantum Hall effect (IQHE) at strong disorder and weak magnetic field is
studied in a lattice model.  A generic sequence by which 
the  IQHE plateaus disappear is revealed: higher IQHE plateaus  always vanish earlier than lower ones, and  extended  levels between those plateaus do not float up in energy but  keep 
merging  together after the destruction of  plateaus.  All of these features remain to be true in the weak-field limit as shown by the thermodynamic-localization-length calculation. Topological characterization in terms of Chern integers provides a simple  physical explanation and suggests a qualitative difference between the lattice and  continuum models. 
\end{abstract} 
 
]

It is an important issue how the integer quantum Hall effect 
vanishes in the weak magnetic field limit. Several years 
ago, Khmel'nitzkii\cite{khm} and Laughlin\cite{laughlin} both argued that 
extended states at the 
centers of Landau levels would not disappear discontinuously nor merge 
together, but should rather float up indefinitely towards higher energy 
to pass the Fermi level. This argument is also a key scenario 
in the global phase diagram proposed by Kivelson, Lee and Zhang\cite{klz} for
the quantum  Hall effect. As a consequence, a direct transition from a higher 
IQHE state to an insulator state is forbiden.\cite{klz} Recently such a floating picture
for extended states has been challenged by a numerical study by Liu {\it et 
al.}\cite{lui} based on a tight binding model (TBM), in which extended levels as
represented by peaks of a finite-size localization length are found to disappear
without changing their position at strong disorder. It indicates that the transition of IQHE 
states to an insulator occurs without involving a floating up of extended 
states. However, Yang and Bhatt\cite{yang} recently argued that this feature in
a lattice model may not be contradictory  to the conventional floating 
picture in the continuum limit. The argument goes as follows. They observed 
that in the TBM the critical strength $W_c$  of disorder by which all extended 
states vanish is roughly $W_c\sim 6t$ ($t$ is the hopping integral) at 
different magnetic flux strengths. If such a {\it finite} $W_c$ indeed persists 
into the weak magnetic-field case,   a weaker disorder strength ($W
< W_c$) will not be able to destroy extended states  and  the only way to get into an
insulator phase in the  weak-field limit  would be a floating up of extended levels 
as suggested by Khmel'nitzkii and Laughlin. 

As these numerical calculations based on the TBM are all performed in 
relatively strong magnetic fields,  one needs more conclusive evidences 
in order to resolve the above controversy and make the case relevant to the 
weak-field limit, which is of experimental interest. In this paper, we reexamine 
this problem in both strong and {\it weak} magnetic fields and  present  convincing
results with regard to the physical mechanism of the disappearance of the IQHE and its
behavior in the weak-field limit.  Firstly,  we show 
a systematic destruction of the IQHE at strong disorder, in which  the IQHE plateaus 
are found to be destroyed in a one-by-one order from high to low energies. During such a 
process,  extended levels separating different IQHE plateaus merge together and then 
disappear without floating  up in energy, and the lowest extended level is the last to vanish 
at a critical disorder strength $W_c$.  Such a picture for 
destroying the IQHE is  shown to persist into the {\it weak field} limit, where 
$W_c$  is found to decrease continuously with
the decrease of magnetic fields, and can be extrapolated to zero at zero field
limit, contrary to a field-independent $W_c$  as suggested in 
Refs.\onlinecite{yang,ando}. A topological  reason for such a destruction of the IQHE by 
disorder is given based on the  Chern number which characterizes extended 
states:\cite{thou2} nonzero Chern integers 
with opposite sign are moving down from the band center to annihilate those in 
the lower-energy extended levels, and eventually a total annihilation of Chern 
numbers leads to a global insulating phase. In contrast to the 
finite-size localization length calculation in Ref. \onlinecite{lui}, though,  we 
find that neighboring extended levels do merge together  before their disappearance,  as the 
result of moving and annihilation of  Chern numbers. 

The TBM Hamiltonian is given as follows: 
\begin{eqnarray*} 
H=-\sum_{<ij> } e^{i a_{ij}}c_i^+c_j + H.c. +\sum _i w_i c^+_i c_i , \nonumber
\end{eqnarray*} 
where the hopping integral $t$ is taken as the unit, and $c_i^+$ is a  fermionic creation 
operator with $<ij>$  referring to two  nearest neighboring sites.  A uniform  
magnetic flux per plaquette is  given as  $\phi=\sum _ {\Box} a_{ij}=2\pi/M$, where the  
summation runs over four links around a plaquette.
And $w_i$ is  a random potential with strength $|w_i|\leq W/2$ (note that the disorder 
strength $W$ here is twice bigger than that defined in Ref.\onlinecite{yang}), and the
white  noise limit is considered with no correlation among different  sites for $w_i$.
 
An overall picture for the Hall conductance calculated by the Kubo
formula is shown in Fig. 1 with the  flux strength $\phi=2\pi/8 $
(only $E<0$  part is shown here).  At weaker disorder  ($W=1$),  three well-defined IQHE plateaus 
at $\sigma_H=\nu e^2/h$  ($\nu=1,2,3$) are clearly shown, corresponding to four Landau levels 
centered at the jumps of the Hall conductance at $E\leq 0$.  With $W$ increasing from  
$1$ to $6$, one sees a systematic destruction of these IQHE plateaus.  For instance, at $W=2$ the
$\nu=3$  plateau is the first beginning to disappear, while at  
$W=3$ and $4$ the $\nu=2$ IQHE plateau starts to vanish too. Meantime  the lowest  plateau still 
remains well defined at $W=4$ and, in the inset of  Fig. 1,  such a $\nu=1$ IQHE plateau region 
is partially shown at different sample sizes, where all the curves 
cross at a  fixed-point corresponding to the Hall conductance $\frac 1 2 e^2/h$  which is to be
extrapolated to a sharp step of jump  in the  thermodynamic limit. By contrast, 
in those higher energy  regions without the plateaus the Hall conductance is seen to monotonically  decrease at larger lattice sizes.  A breakdown of the lowest  IQHE plateau is eventually found at larger 
$W$'s ($W=6$ case is shown in Fig. 1). Hall conductances at
weaker flux strengths with $M=11$, $16$ and $24$ all exhibit the similar  generic
features as in Fig. 1.  
 
A key thing to understand the above evolution of IQHE plateaus is the  
localization-delocalization transition. It is a well-known fact that in the IQHE 
an extended state can be characterized by a topological index,  
namely, a non-zero Chern integer.\cite{thou2} And the boundary-condition-averaged Hall conductance 
is a summation of all those Chern numbers carried by states below  
Fermi surface:\cite{thou2} $<\sigma _H (E_f)>=e^2/h \sum_{\varepsilon_m < E_f} C^{(m)}$, 
where the Chern number $C^{(m)}$ is always an integer. Since its distribution in 
the thermodynamic limit decides delocalization regions, one may define  
states with nonzero Chern integers as extended states and calculate the corresponding
density of extended states $\rho_{ext}$.\cite{huo} The results are presented in  
Fig. 2 with the same flux strength as in Fig. 1. At $W=1$ (weak disorder), well-defined 
peaks of  $\rho_{ext}$ are located at centers of Landau-level bands, separated by localized  
regions represented by plateaus in Fig. 1. Widths of these 
peaks will approach to zero in the thermodynamic limit.\cite{huo} Total Chern 
number for each of three lower-energy peaks is found to be exactly $+1$, which is the 
reason leading to three quantized Hall plateaus at $+1$, $+2$, and $+3$ in unit
of $e^2/h$ shown in Fig. 1, when the Fermi surface is located between these 
peaks. The  last peak closest to the band center carries a total
 Chern number $-3$, which guarantees that the Hall conductance in Fig. 1  
falls back to zero beyond the third plateau when the Fermi energy approaches 
$E=0$.  This is a peculiar feature of a lattice model since when the whole band is  
half-filled, the Hall conductance has to be zero. These Chern integers are conserved as long as they are
well separated by localization regions, just like the quantization of the plateaus themselves. 
 
The distribution of Chern integers (thus the extended states) begins to qualitatively change at stronger disorders. At $W=2$ and  $3$, two higher-energy peaks in Fig. 2  start to merge due to 
disorder scattering. Because  these two peaks carry total Chern numbers of $+1$ and $-3$, respectively,  
a mixing and  annihilation of these Chern numbers results in a 
disappearance of the third plateau in Fig. 1, and leads to a substantial 
reduction of the magnitude of the Hall conductance in that region. By contrast,  the lowest-energy  peak remains separated from the rest spectrum up to $W=5$. In the insert of Fig. 2,  such a  peak at $W=4$ as located at the  energy $E_{c1}\simeq -3.4$ is shown to become
narrower and sharper with the increase of lattice sizes, which indicates  that a well-defined mobility  
edge still exists at $E_{c1}$. Such a peak disappears finally as the negative Chern numbers reach to the region around $E_{c1}$ as shown at $W=6$, and we find a monotonic decrease of $\rho_{ext}$ with 
larger lattice sizes which corresponds to a localization in the whole region.  
Therefore,  the destruction of the IQHE at strong magnetic fields in the TBM can be well understood physically as due to a moving down of negative Chern integers from  band center,  caused by strong disorders,  which mix with the positive Chern integers located at lower-energy extended levels and eventually annihilate them, leading to the insulator transition of the whole system. 

Two questions remain to be answered here:  Firstly, since only a finite-size analysis has been done so far, one may wonder if the above picture of the insulator transition is still  to be true in the thermodynamic limit, and secondly,  whether  such a picture also persists continuously into  weak magnetic field case which is more relevant to the experimental situation. 
To answer the first question, we need to determine the thermodynamic localization length by a standard finite-size scaling method.\cite{note,Mac} The results for the same field strength as in Figs. 1 and 2 are plotted in Fig. 3 for the case of $W= 4$ and $5$. At $W=4$,  the thermodynamic  localization length follows a scaling behavior $\xi\sim  |E-E_{c1}|^{-x}$ with the exponent $x = 2.4\pm 0.1$ on the two 
sides of the critical  energy $E_{c1}$, which verifies the existence of a delocalization fixed-point shown 
in the  inset of Fig. 1 (the extended-state position $E_{c1}=-3.40$ also coincides with the one previously determined by the  Chern number calculations). The  localization length $\xi$ exhibits an another divergence near $E_{c2}\simeq -2.52$ ($\rho_{ext}$ in the insert of Fig. 2 also shows a second peak 
here). Above $E_{c2}$ one finds the localization length to be always finite up to the band center, indicating a localized region with zero Hall conductance as the negative Chern integers move down to  $E_{c2}$.  Thus, between $E_{c1}$ and $E_{c2}$ there must remain a Hall conductance plateau as shown in Fig. 1. With  $W$ increasing from  $4$ to $5$,  $E_{c1}$ is only slightly reduced (from $-3.40$ to $-3.48$) whereas $E_{c2}$ moves down much more quickly (from $-2.52$ to $-3.0$), such that the $\nu=1$ IQHE plateau between $E_{c1}$ and $E_{c2}$ becomes 
narrower, and eventually  these last two extended levels merge and disappear at a critical $W_c$.  Correspondingly the $\nu=1$ plateau vanishes in Fig. 1 beyond this point.  Thus, in contrast to the finite-size localization length calculations in Ref.\onlinecite{lui} where all the peaks of the localization length remain unchanged before their disappearance, the present thermodynamic localization-length  data clearly indicate a moving-down and mixing of  extended levels before they disappear, consistent with the  Chern number interpretation for the transition.

The magnetic flux strength $\phi=2\pi/M$  at $M=8-24$ used in the previous calculations are relatively high fields. In order to answer the second question about the weak field limit, we have to reduce the field strength substantially. In the following we inspect  this case by examining the critical disorder $W_c$. Note that by increasing $W$, we always find that higher-energy extended levels disappear first, while the lowest one vanishes in the last without floating up in energy (which actually  slightly shifts downwards).
Then by following the evolution of the lowest 
extended level which can be determined from the thermodynamic localization length, one can decide a critical $W_c$, beyond which no extended states exist in the whole energy regime. $W_c$'s obtained in 
this way are shown in Fig. 4 versus $1/M$,  the magnetic flux strength in unit of $2\pi$. We have been able to reach a weak field as low as
$\phi= {2\pi}/{384}$ or $M=384$,  and Fig. 4 shows that $W_c$  monotonically decreases with the magnetic field strength which is extrapolated to zero at zero magnetic field.  This result is at variance with the conjecture made by Yang and Bhatt\cite{yang} about roughly a constant $W_c$ ($\sim 6$) at $M\rightarrow \infty$ as extrapolated from the data within $M<13$. On the contrary, $W_c$ in Fig. 4 is
 well fit by a $\frac{1}{\sqrt{M}}$ law at $M\geq 16$ which in turn means $W_c\propto \sqrt{\omega_c}$
($\omega_c$ is the magnetic cyclotron energy and all quantities here are dimensionless as scaled by $t$ which is taken as the unit). If one estimates the Landau level broadening at weak field as $\Gamma \sim \frac{W^2}{t}$,\cite{lui} then $\Gamma\propto \omega_c$ is found at the critical $W_c$. It implies that once the neighboring Landau levels are coupled together due to the disorder scattering, the negative Chern integers near the band center will be able to flow all the way down to the band bottom and lead to the insulator transition.  In other words, it is not necessary to have  a strong $W_c$ {\it comparable} to the bandwidth in order to mix all the positive and negative  Chern numbers. 
This is  significant because the weak-field limit in {\it real materials} corresponds to the case that the Landau level spacing is much less than the bandwidth, and our results reveal that even in this limit  the lattice effect (due to which one gets negative Chern numbers near the band center) is still crucial as far as the metal-insulator transitions are concerned.  It suggests that  the continuum model is  {\it oversimplified} in describing such phenomena in real systems where the lattice structure is always present. 

Finally, we would like to make several remarks. Firstly, a limited floating-up picture can be still seen  if the Landau-level {\it filling number} is fixed. The reason is that the number of localized states below, say, the lowest extended level at $E_{c1}$ can be {\it enhanced} by stronger disorders, so that when the filling number is fixed the Fermi level  actually moves down, or, relatively the extended level shifts up,  as already pointed out by Yang and Bhatt.\cite{yang}  But such a relative ``floating up'' for the extended levels before their destruction is  very limited and only within an order of one Landau level spacing, certainly not infinite, for the lowest-energy extended level as checked by us up to $M=192$. Secondly, as with the earlier work,\cite{lui} the present results suggest that a direct transition from higher IQHE to a zero-Hall-effect insulator becomes possible,  after the IQHE plateau structure in Fig. 1 collapses from high energy downwards to the Fermi!
 surface regime  at a sufficient 
global phase diagram.  We shall leave discussions about  comparison with experimental measurements elsewhere. Lastly, a further theoretical understanding is still needed about why the negative Chern numbers near the band center can so {\it easily} move down to the band bottom once the neighboring Landau levels start to mix by disorder scattering.
 
{\bf Acknowledgments} -The authors would like to thank T. Xiang, X. C. Xie, and K. Yang for
stimulating and helpful  discussions. The 
present work is supported  by TCSUH, and TARP under a grant no.\# 3652182, and a grant from Robert
Welch foundation.

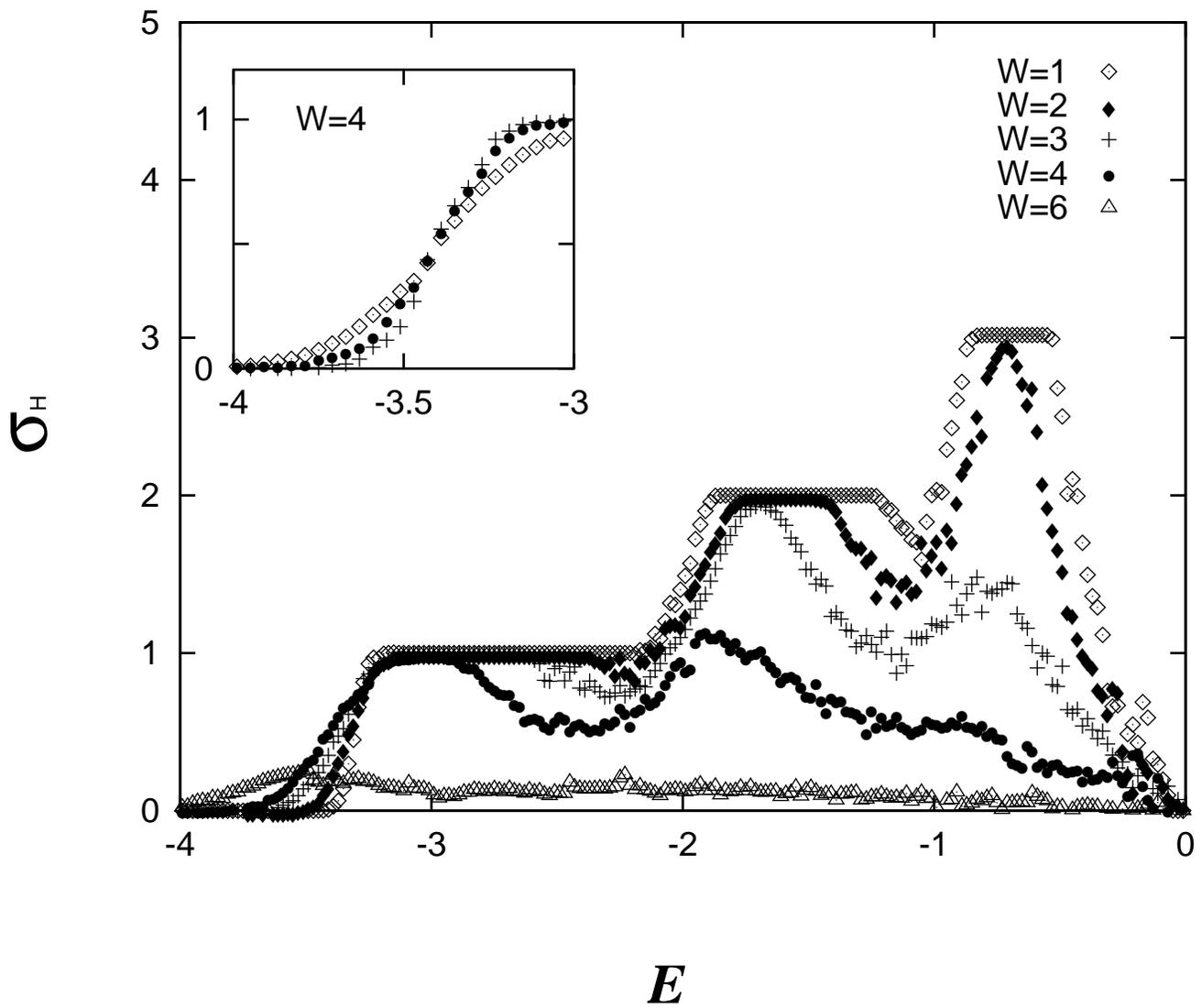
Fig. 1.  The Hall conductance $\sigma _H$ as a function of energy E
is plotted for different disorder strength  $W$'s, at a lattice size 
$16 \times 16$. The inset shows the evolution of the $\nu=1$ IQHE 
plateau at $W=4$ with lattice size varying form $8\times 8$ ($\diamond $), 
$16\times 16$ ($\bullet $), to  $24\times 24$ ($+$).

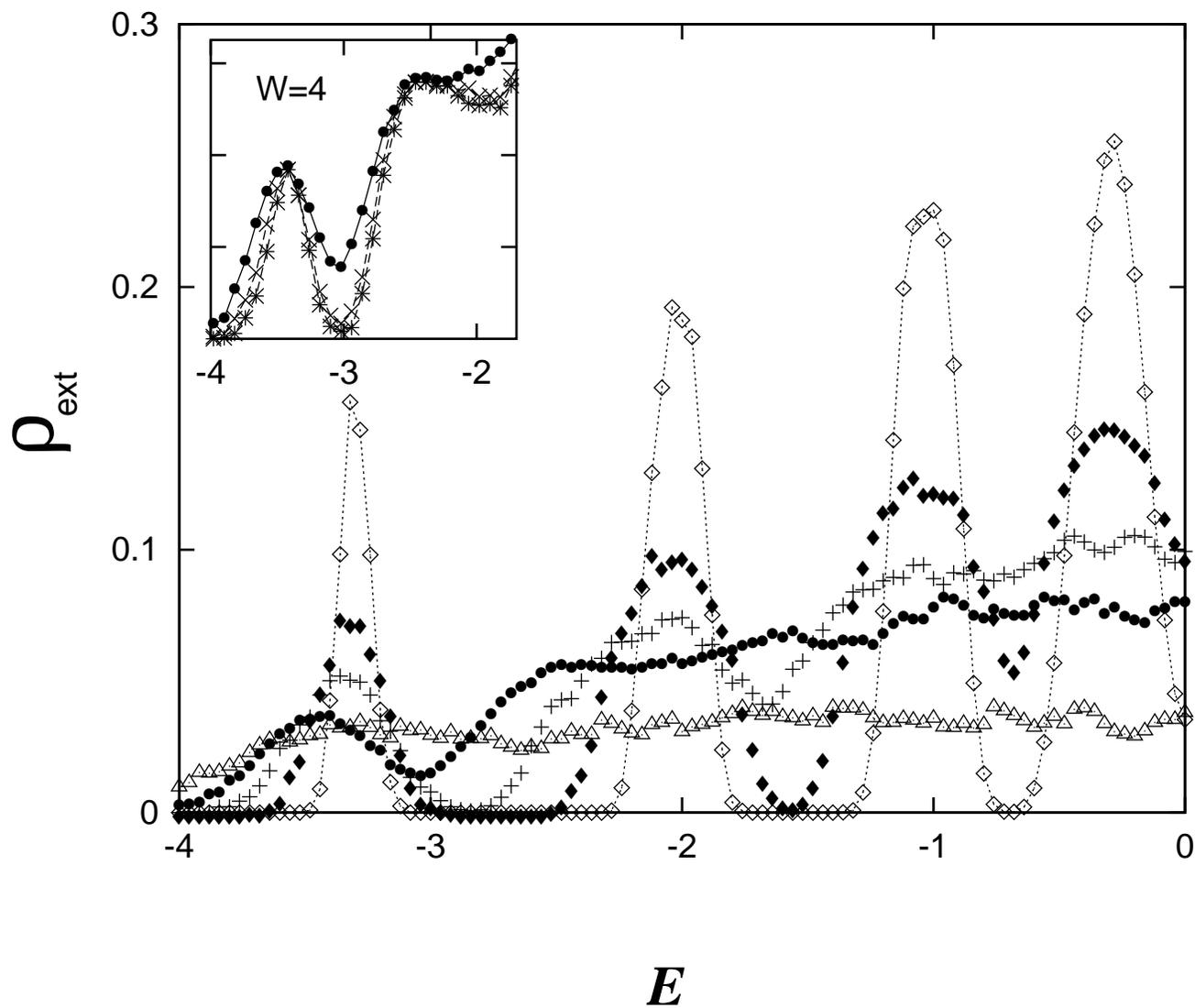
Fig. 2. The density of extended states $\rho_{ext}$ versus $E$. Disorder 
strength $W$'s  are the same as in 
Fig. 1  with a lattice size $8 \times 8$ (about 400 random-potential 
configurations are used). The inset shows $\rho_{ext}$ at
$W=4$ with lattice sizes varying as $8\times 8$ ($\bullet$), $16\times 16$
($\times$), and $24\times 24$ ($\ast$).
          
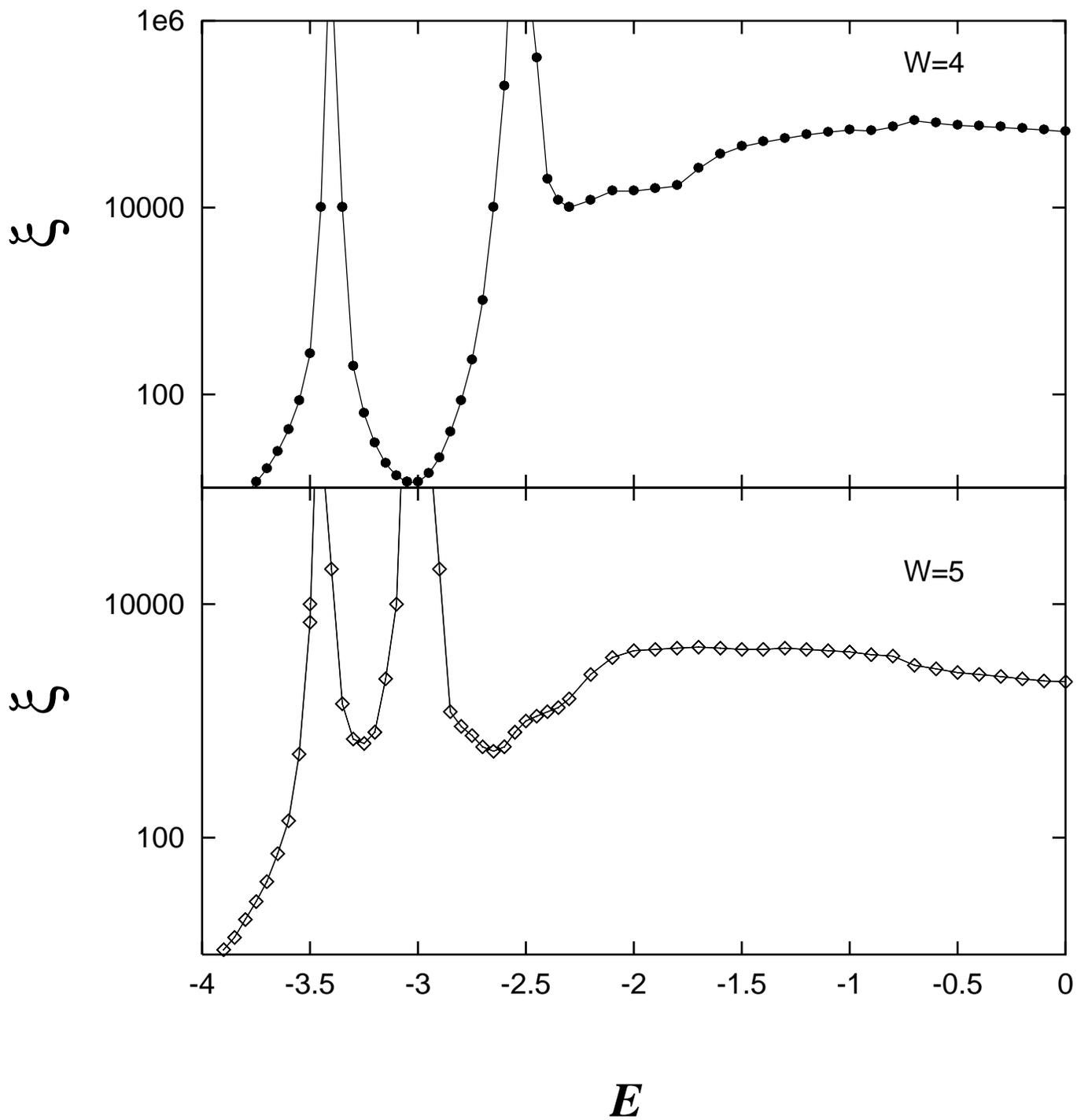
Fig. 3. Thermodynamic localization length  $\xi$ is shown as a function of energy E at $W=4$ and
$5$, respectively.  Critical behaviors  near divergent points are fit  by scaling laws (see text).            
 
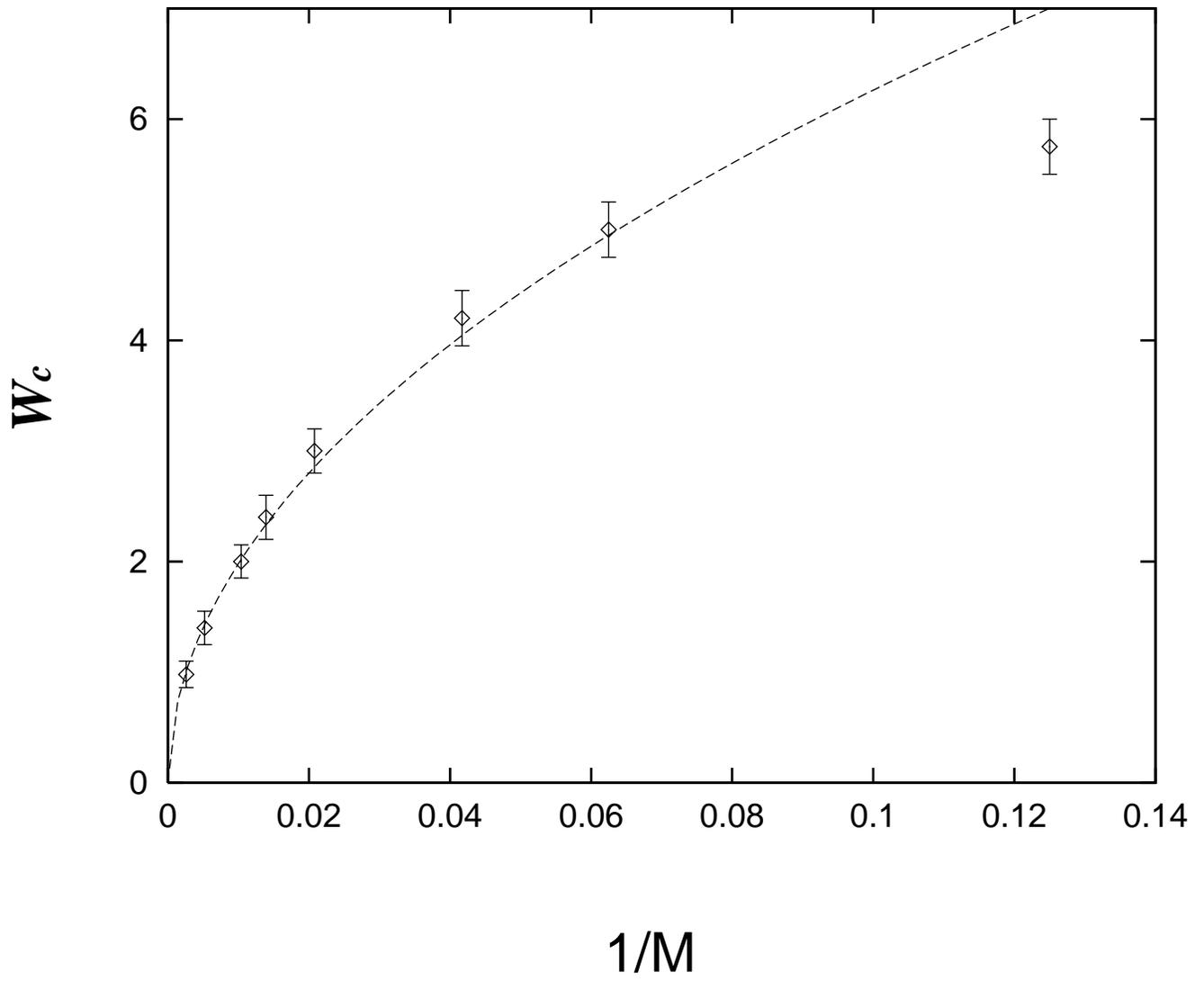
Fig. 4.  The critical strength of disorder $W_c$ versus the magnetic flux strength (in unit $2\pi$)
 ${1}/M$. The dashed curve of a $1/\sqrt{M}$ law well fits the data  at $M\geq 16$. 


\begin{references}
\bibitem{note} The method outlined in Ref. \onlinecite{Mac} is employed to determine the finite-size 
localization length based on strip lattices with the width $L$ varying from $16 $ to $80$ and the length over  $10^5$. Then the thermodynamic localization length $\xi$ can be extracted in terms of a standard one-parameter scaling analysis.
\end{references}
\end{document}